 \documentclass[final,1p,times]{elsarticle}

\usepackage{amssymb}
\usepackage{amsmath}

\usepackage[colorlinks=true,linkcolor=blue,citecolor=blue,urlcolor=blue]{hyperref}
\usepackage[normalem]{ulem}
\usepackage[table,xcdraw]{xcolor}

\begin{document}

\begin{frontmatter}



\title{Generalized flux trajectories:\\ New insights into partially coherent Airy beams}


\author{\'Angel S. Sanz and Rosario Mart{\'\i}nez-Herrero}

\affiliation{organization={Department of Optics, Faculty of Physical Sciences,
		Universidad Complutense de Madrid},
            addressline={Pza.\ Ciencias 1}, 
            city={Madrid},
            postcode={28040}, 
            country={Spain}}

\begin{abstract}
The propagation of Airy beams in free space is characterized by being non dispersive, which warrants the shape invariance of their intensity distribution, and self-accelerating along the transverse direction.
These distinctive traits are still present in partially coherent Airy beams as long as the reach of their back tail (and hence their energy content) is not importantly reduced.
To investigate the effects associated with the decrease of the beam coherence and its power content (by smoothly reducing the reach of their back tails), here we introduce a novel and insightful methodology based on a generalization of the concept of flux trajectory for paraxial partially coherent beams.
This methodologies emphasizes the role of phase relations, thus helping to clarify why and how the beam smears out spatially along its propagation.
This formalism, though, is general enough to tackle other types of structured light beams with whatever degree of partial coherence, from full coherence to total incoherence.
\end{abstract}




\end{frontmatter}




\section{Introduction}
\label{sec1}

Over the last ten years, the field of structured light has received much attention, experiencing a remarkable growth due to its promising technological applications \cite{forbes:JOpt:2016,andrews-bk:2008,forbes:AVSQuantumSci:2019,forbes:JOpt:2018,forbes:LaserPhotRev:2019,forbes:NatPhotonics:2021,forbes:LSA:2022}.
This has provoked the implementation and development of novel experimental methods aimed at designing, generating, and optimizing new types of structured light beams, as well as analytical approaches to examine and get a deeper understanding of their properties and physical behavior.
In this regard, it is worth mentioning the family of self-accelerating beams, which has also made considerable advances in terms of both theory and experiment, and also in connection to their applications (see Ref.~\cite{christodoulides:Optica:2019}, and references therein).

The self-accelerating propagation and shape-invariance properties inherently linked to Airy beams, as a paradigmatic representative of this family, have been regarded to be a consequence of their coherence \cite{berry:AJP:1979}.
Recently, though, it has been noticed \cite{sanz:JOSAA:2022} that such a behavior can also be related to a push of the infinite tail on the rearmost parts of the beam, which, under the lack of an opposed contribution, undergo the transverse self-acceleration that characterizes them.
This phenomenon vanishes as soon as the beam is truncated, since the (infinite) boost received from the infinite tail disappears; in this case, the beam starts spreading backwards along the transverse direction (but not along the propagation one), which eventually leads to the also disappearance of the spatial self-invariance of the corresponding intensity distribution.
This is precisely what happens with realistic Airy beams generated in the laboratory \cite{christodoulides:PRL:2007,christodoulides:OptLett:2007}.
On the contrary, it has also been observed \cite{sanz:PRA:2022} that the properties displayed by ideal Airy beams can still be preserved under total absence of coherence provided the infinite back tail of the beam remains, i.e., under infinite energy conditions.
This type of structure light beams have recently been applied in the transmission of images by means of arrays of Airy beams \cite{Zhou:ApplOpt:2012}, imaging out of paraxiality conditions \cite{orlov:LSA:2022}, or the fabrication of optical needles \cite{orlov:OLT:2024}.
A wide perspective on different applications of this type of structured light beams are discussed can be found in Ref.~\cite{christodoulides:Optica:2019}.

In order to investigate and explain the propagation properties of partially coherent beams, such as partially coherent Airy beams, here we present a generalized theoretical framework, which relies on the quantum concept of Bohmian trajectory, brought in to the partially coherent structured light field.
In quantum mechanics, these trajectories arise in a natural manner after integrating in time the velocity field resulting from the ratio between the quantum current density and the probability density, thus describing the transport of probability in the position representation \cite{bohm:PR:1952-1,sanz-bk-1}.
In the same way, in paraxial optics, analogous trajectories arise for the transport of energy from the effective velocity that results from dividing the flux vector by the intensity distribution, although the parameterization is not in terms of time, but of the coordinate accounting for the propagation direction \cite{sanz-bk-1,sanz:JOSAA:2012,sanz:ApplSci:2020}.
This concept, which has already been considered in the study of the propagation properties of different types of coherent structured light beams \cite{sanz:ApplSci:2020,sanz:JOSAA:2022,sanz:PLA:2024,sanz:OE:2024}, relies on a novel definition of effective transverse velocity, formally grounded on a general representation of the cross-spectral density (CSD).
This thus set up a direct analogy to its non-relativistic quantum partner in terms of the reduced density matrix \cite{sanz:EPJD:2007}.
Accordingly, the trajectories arising from the integration of the corresponding propagation equation of motion along the corresponding propagation coordinate describe how the beam energy distributes locally, along the transverse direction, as the beam moves forward.
This novel methodology thus introduces a substantial change in the way to conceive and understand the propagation of partially coherent light, since it provides us with a synergistic combination of the usual analysis of terms of intensity distributions, on the one hand, with an alternative representation of the process based on the transverse energy flux.
The spatial changes arising from the latter can then be visualized, at a local level, in terms of trajectories, or at a global level, in terms of the associated effective velocity field, which is a measure of local phase changes.

Specifically, this novel definition of effective transverse velocity is used to analyze the properties displayed by partially coherent Airy beams defined in terms of the corresponding CSDs, as introduced in Ref.~\cite{sanz:PRA:2022}, where their propagation is mainly determined by two competing, intertwined mechanisms: energy finiteness and coherence. As is shown, two different regimes can be distinguished in terms of which one of those mechanisms prevails over the other, which finds an interesting interpretation in terms of the corresponding effective transverse velocity. Furthermore, the features displayed by the associated ensembles of trajectories provide us with valuable information about the beam local development at later stages of the propagation. All this information is then considered to study the propagation of partially coherent superpositions of two overlapping Airy beams, where we can notice again the persistence of those two clearly different regimes.

The work is organized as follows.
A detailed account of the theoretical framework is provided in Sec.~\ref{sec2}, which allows us to introduce the definition of the generalized trajectories here presented as well as the specific theoretical aspects involved in the propagation of partially coherent Airy beams and their superpositions.
In Sec.~\ref{sec3}, a series of illustrative results from numerical simulations are presented and discussed for three scenarios with partially coherent Airy beams and also with the corresponding superpositions.
Specifically, we consider quasi-infinite energy beams with a rather poor coherence, finite energy beams with a good coherence, and an intermediate case.
Finally, a series of summarizing remarks are discussed in Sec.~\ref{sec4}.


\section{Theory}
\label{sec2}


\subsection{Generalized trajectory framework}
\label{sec21}

Consider a paraxial monochromatic beam described by the general expression
\begin{equation}
\Psi ({\bf r}_\perp,z) = \psi({\bf r}_\perp,z) e^{ikz}
\label{eq1}
\end{equation}
where ${\bf r}_\perp = (x,y)$ stands for the transverse coordinate vector, and $k = 2\pi/\lambda$ (with $\lambda = \lambda_0/n$ and $\lambda_0 = c/\omega$), and the complex amplitude $\psi({\bf r}_\perp,z)$ satisfies Helmholtz's paraxial equation,
\begin{equation}
i\ \frac{\partial \psi}{\partial z} = - \frac{1}{2k} \nabla_\perp^2 \psi .
\label{eq2}
\end{equation}
On the right-hand side of the latter equation, $\nabla_\perp^2 \equiv \partial^2/\partial x^2 + \partial^2/\partial y^2$ denotes the transverse Laplacian.
If Eq.~(\ref{eq2}) is multiplied by $\psi^*$ on both sides and then, to the resulting equation, we subtract its complex conjugate version, we obtain
\begin{equation}
\frac{\partial |\psi ({\bf r}_\perp,t)|^2}{\partial z} =
- \frac{1}{2ik} \left[ \psi^* \left( \nabla_\perp^2 \psi \right)
-  \left( \nabla_\perp^2 \psi^* \right) \psi \right] ,
\label{eq3}
\end{equation}
The right-hand side in this equation can still be further simplified, which leads to the conservation or continuity equation for the light intensity distribution,
\begin{equation}
\frac{\partial I}{\partial z} = - \nabla_\perp \cdot \left[ I\
\frac{1}{k}\ {\rm Re} \left( \frac{-i\nabla_\perp \psi}{\psi} \right) \right] ,
\label{eq4}
\end{equation}
if we multiply on both sides by the corresponding constant prefactors, since $I({\bf r}_\perp,t) \propto |\psi ({\bf r}_\perp,t)|^2$, which does not affect the term still depending on the beam amplitude $\psi$ on the right-hand side.
Since $|\psi|^2$ is proportional to the energy density distribution, the term between square brackets on the right-hand side can be defined as the transverse flux vector,
\begin{equation}
{\bf j}({\bf r}_\perp,z)
= I({\bf r}_\perp,z) {\bf v}_\perp ({\bf r}_\perp,z) ,
\label{eq5}
\end{equation}
where the vector field
\begin{equation}
{\bf v}_\perp ({\bf r}_\perp,z) = \frac{{\bf j}({\bf r}_\perp,z)}{I({\bf r}_\perp,z)}
= \frac{1}{k}\ {\rm Re} \left( \frac{-i\nabla_\perp \psi}{\psi} \right) .
\label{eq6}
\end{equation}
specifies the local rate of change of the flux at the point ${\bf r}$, and hence it behaves in analogous way to a local velocity field.
Accordingly, we shall refer to this quantity as a transverse effective velocity field.

It is worth noting that, if the complex field amplitude is recast in polar form, i.e.,
\begin{equation}
\psi({\bf r}_\perp,z) = A({\bf r}_\perp,r) e^{iS({\bf r}_\perp,z)} ,
\label{eq7}
\end{equation}
we can extract valuable information from the effective velocity field (\ref{eq6}), as it reads as
\begin{equation}
{\bf v}_\perp ({\bf r}_\perp,z) = \frac{1}{k}\ \nabla_\perp S({\bf r}_\perp,z) .
\label{eq8}
\end{equation}
This expression explicitly makes evident that any change in the flux (and hence the light intensity distribution) is directly connected with the local variations undergone by the phase of the beam along the transverse direction during the propagation.
Hence, to understand at a deeper level the implications of such spatial variations, we can proceed as in classical mechanics and integrate (along the propagation coordinate, $z$) the effective equation of motion
\begin{equation}
\frac{d {\bf r}_\perp}{dz} \equiv {\bf v}_\perp({\bf r}_\perp,z) .
\label{eq9}
\end{equation}
The trajectories arising from this equation of motion will thus describe any local effect of the phase on the distribution of light intensity.
Actually, if the set of initial conditions is chosen in such a way that they distribute along the transverse direction in a given plane (e.g., the input plane) according to the intensity distribution in such a plane, their distribution in a different transverse plane will be in agreement with the intensity distribution in such a plane.
This thus settles the grounds for a methodology aimed at analyzing the propagation of structured-light scalar beam propagation under paraxial conditions on an event-by-event basis \cite{sanz:JOSAA:2022}.

The above procedure to reach Eq.~(\ref{eq9}) from the paraxial Helmholtz equation is analogous to the process followed to obtain the quantum hydrodynamic or Bohmian trajectories from Schr\"odinger's equation associated with a matter wave in quantum mechanics \cite{sanz-bk-1}.
Nonetheless, it is also possible to reach the same result considering the vector nature of the beam and then assuming paraxiality conditions \cite{allen:PRA:1992,sanz:ApplSci:2020}.
It is worth noting that, in this case, after averaging in time the Poynting vector, two contributions are well distinguished, namely, one simply related to the forward paraxial propagation and another associated with the transverse local phase variations undergone by the beam.
This convenient decoupling of the Poynting vector, between longitudinal and transverse contributions, has been considered precisely by Broky {\it et al.}~\cite{broky:OE:2008} to investigate the self-healing properties of Airy beams.

Equation~(\ref{eq9}) is only valid for fully coherent monochromatic paraxial beams.
However, we show next a route to generalize it to quasimonochromatic paraxial partially coherent beams.
To do so, we consider as the main descriptor of the beam its cross spectral density (CSD), $W({\bf r}'_\perp,{\bf r}_\perp,z)$.
Physically, this quantity specifies the field correlations between two spatial points at a given plane \cite{mandelwolf-bk}.
Specifically, making use of the well-known result \cite{charo:NuovCimB:1979,mandelwolf-bk} that any CSD can be recast in terms of a set of field modes, $\{ \phi_n \}$,
\begin{equation}
W({\bf r}'_\perp,{\bf r}_\perp,z) =
\sum_n p_n \phi_n^* ({\bf r}'_\perp,z) \phi_n ({\bf r}_\perp,z) ,
\label{eq13}
\end{equation}
where the $p_n$ denote the positive weight associated with the field mode $\phi_n$, which satisfies Helmholtz's paraxial equation,
\begin{equation}
i\ \frac{\partial \phi_n ({\bf r}_\perp, z)}{\partial z} = - \frac{1}{2k} \nabla_\perp^2 \phi_n ({\bf r}_\perp, z) .
\label{paraxfn}
\end{equation}
According to the above condition, the field modes will also satisfy a continuity equation,
\begin{equation}
\frac{\partial |\phi_n ({\bf r}_\perp,z)|^2}{\partial z} = - \nabla_\perp \cdot {\bf j}_n ({\bf r}_\perp,z) ,
\label{eqcont}
\end{equation}
where ${\bf j}_n$ plays the role of a (transverse) flux vector, defined as
\begin{equation}
{\bf j}_n ({\bf r}_\perp,z) = \frac{1}{2ik} \left\{ \phi_n^*({\bf r}_\perp, z) \left[ \nabla_\perp \phi_n ({\bf r}_\perp, z) \right]
- \left[ \nabla_\perp \phi_n^* ({\bf r}_\perp, z) \right] \phi_n ({\bf r}_\perp, z) \right\} .
\label{eqjn}
\end{equation}
Note that, in this context, the fully coherent scenario corresponds to a single term on the right-hand side of Eq.~(\ref{eq13}); that is, the coherent case is equivalent to specifying the CDS in terms of a one single field mode.

Note that, within this formulation, the transverse light intensity distribution for the partially coherent beam is given by the diagonal elements of the CSD~(\ref{eq13}), i.e.,
\begin{equation}
\bar{I}({\bf r}_\perp, z) \equiv W({\bf r}'_\perp,{\bf r}_\perp,z)\big\vert_{{\bf r}'_\perp = {\bf r}_\perp} = \sum_n p_n |\phi_n ({\bf r}_\perp,z)|^2 .
\label{transint}
\end{equation}
On the other hand, considering the sum over modes (with the corresponding weights $p_n$) on the left-hand side of Eq.~(\ref{eqcont}), we readily find that the variation of the intensity profile (\ref{transint}) along the propagation direction also follows a continuity equation, which reads as
\begin{equation}
\frac{\partial \bar{I} ({\bf r}_\perp, z)}{\partial z} = - \nabla_\perp \cdot \bar{\bf j}({\bf r}_\perp,z) ,
\label{intderiv}
\end{equation}
where we introduce the definition of the partially coherence transverse flux, as
\begin{equation}
\bar{\bf j}({\bf r}_\perp,z) \equiv \sum_n p_n\ {\bf j}_n({\bf r}_\perp,z) ,
\label{eqj}
\end{equation}
that is, the averaged flux associated with the CSD~(\ref{eq13}) results from an incoherent superposition of the transverse fluxes associated with the field modes that contribute to the CSD, with the same weights.
More importantly, concerning the approach that is being developed here, the fact that the light intensity distribution associated with the partially coherent beam is directly related to an averaged flux through a neat continuity equation, Eq.~(\ref{intderiv}), allows us to establish a direct link between both analogous to Eq.~(\ref{eq5}), and hence to define a generalized transverse effective velocity field for a partially coherent beam, as
\begin{equation}
\bar{\bf v}_\perp({\bf r}_\perp,z) = \frac{\bar{\bf j}({\bf r}_\perp,z)}{\bar{I}({\bf r}_\perp,z)} .
\label{eq10}
\end{equation}
Physically, this effective velocity is the overall driving field that makes the intensity distribution of the partially coherent beam to disperse spatially along the transverse coordinate in one way or another, as the beam moves forward, depending on the degree of coherence of the beam.

Taking into account the above findings, we may wonder about the expression of the transverse velocity field for the partially coherent beam when the latter is assumed to be a collection of $M$ random measures of the beam, $\psi: \{\psi_m,\ {\rm for}\ m=1\ {\rm to}\ M\}$, instead of orthogonal field modes, where each realization $\psi_m$ is independent.
Without loss of generality, we assume no bias in the measure of these realizations, so that all of them contribute equally to the CSD, which, in this case, will read as
\begin{equation}
W({\bf r}'_\perp,{\bf r}_\perp,z) = \langle \psi^* ({\bf r}'_\perp,z) \psi ({\bf r}_\perp,z) \rangle
= \lim_{M \to \infty}\ \frac{1}{M} \sum_{m,m'=1}^M \psi_{m'}^* ({\bf r}'_\perp,z) \psi_m ({\bf r}_\perp,z) ,
\label{eq12}
\end{equation}
i.e., in terms of an average over beam realizations.
Note that this description somehow resembles the description of an open quantum system in terms of its reduced density matrix within the quantum state diffusion picture \cite{diosi:JPA:1988,diosi:PLA:1988,gisin:JPA:1992,gisin:JPA:1993,breuer-bk:2002,percival-bk}, where the loss of coherence arises from the statistical averaging over realizations of the state system vector states (or, if projected onto a position basis set, the associated field amplitudes).
Now, taking into account the decomposition (\ref{eq13}) for the CSD, we assume that the beam amplitudes $\psi_m$ can be recast as linear combinations of field modes $\phi_n$, as
\begin{equation}
\psi_m({\bf r}_\perp,z) = \sum_n c_n^{(m)} \phi_n({\bf r}_\perp,z) .
\label{eq14}
\end{equation}
Specifically, this is possible whenever the condition
\begin{equation}
\lim_{M \to \infty}\ \frac{1}{M} \sum_{m,m'=1}^M c_{n'}^{(m')*} c_n^{(m)} \sim
\delta_{nn'} \left( \lim_{M \to \infty}\ \frac{1}{M} \sum_{m,m'=1}^M c_{n'}^{(m'),*} c_n^{(m)} \right) = \delta_{nn'} p_n .
\label{eq15}
\end{equation}
is satisfied, which results from substituting Eq.~(\ref{eq14}) into Eq.~(\ref{eq12}).
In the intermediate step, we have assumed that, unless $n' = n$, the averaging cancels out, as we are dealing with independent random realizations.
With this choice, the transverse intensity distribution reads as
\begin{equation}
\bar{I} ({\bf r}_\perp, z) = W({\bf r}'_\perp,{\bf r}_\perp,z)\big\vert_{{\bf r}'_\perp = {\bf r}_\perp} = \langle |\psi({\bf r}_\perp,z)|^2 \rangle .
\label{transint2}
\end{equation}
As before, if we calculate the variation of the intensity distribution, specified this time by Eq.~(\ref{transint2}), with $z$, but making use of functional form (\ref{eq12}) for the CSD (with ${\bf r}'_\perp = {\bf r}_\perp$), we obtain
\begin{equation}
\frac{\partial \bar{I} ({\bf r}_\perp,z)}{\partial z} =
\sum_n p_n\ \frac{\partial |\phi_n({\bf r}_\perp,z)|^2}{\partial z} ,
\label{intderiv2}
\end{equation}
again, by virtue of Eq.~(\ref{intderiv}), leads to
\begin{equation}
\frac{\partial \bar{I} ({\bf r}_\perp,z)}{\partial z} =
- \nabla_\perp \cdot \bar{\bf j} ({\bf r}_\perp,z) .
\label{eqcont2}
\end{equation}
%
%
%
Taking into account the explicit expression for the partial contributions ${\bf j}_n$, given by Eq.~(\ref{eqjn}), we note that the variations along $z$ undergone by the intensity distribution, $\bar{I}({\bf r}_\perp,z)$, when recast as an average over realizations of partial intensity distributions, correspond to the generalized flux $\bar{\bf j}({\bf r}_\perp,z)$ introduced through the definition (\ref{eqj}).
This makes the above statistical interpretation in terms of random field realizations more apparent.
Furthermore, we also note that the generalized flux can readily be recast in terms of the CSD, Eq.~(\ref{eq12}), as follows
\begin{eqnarray}
\bar{\bf j}({\bf r}_\perp,z) & = & \frac{1}{2ik} \sum_n p_n \Big\{ \phi_n^* ({\bf r}'_\perp,z) \big[\nabla_\perp \phi_n ({\bf r}_\perp,z)\big]
- \big[\nabla'_\perp \phi_n^*({\bf r}'_\perp,z)\big] \phi_n ({\bf r}_\perp,z) \Big\}_{{\bf r}'_\perp = {\bf r}_\perp}
\nonumber \\
 & = &  \frac{1}{2ik} \Big[ \nabla_\perp W ({\bf r}'_\perp,{\bf r}_\perp,z)
- \nabla'_\perp W({\bf r}'_\perp,{\bf r}_\perp,z) \Big]_{{\bf r}'_\perp = {\bf r}_\perp} ,
\label{eqfluxgen}
\end{eqnarray}
where $\nabla'_\perp$ refers to computing the Laplacian with
respect to the variable ${\bf r}'_\perp$.
From Eq.~(\ref{eq12}) and the last expression on the right-hand side of Eq.~(\ref{eqfluxgen}),
we note that the partially coherent transverse flux can be written, in a more convenient
manner, in terms of the field amplitude $\psi({\bf r}_\perp,z)$ as
\begin{equation}
\bar{\bf j}({\bf r}_\perp,z) =
\frac{1}{2ik} \left\langle \psi^*({\bf r}'_\perp,z) \big[ \nabla_\perp \psi({\bf r}_\perp,z) \big] - \big[ \nabla'_\perp \psi^*({\bf r}'_\perp,z) \big] \psi({\bf r}_\perp,z) \right\rangle
\Big\vert_{{\bf r}'_\perp = {\bf r}_\perp} .
\label{eq17}
\end{equation}

Taking into account the expression for the intensity distribution and the transverse flux, Eqs.~(\ref{intderiv2}) and (\ref{eq17}), the generalized transverse equation of motion (\ref{transint2}), in terms of the random realizations $\psi$, takes the form
\begin{equation}
\frac{d {\bf r}_\perp}{dz} = \frac{1}{2ik} \frac{\left\langle \psi^*({\bf r}'_\perp,z) \big[ \nabla_\perp \psi({\bf r}_\perp,z) \big] - \big[ \nabla'_\perp \psi^*({\bf r}'_\perp,z) \big] \psi({\bf r}_\perp,z) \right\rangle \Big\vert_{{\bf r}'_\perp = {\bf r}_\perp}}{\langle |\psi({\bf r}_\perp,z)|^2 \rangle} .
\label{eq10-xx}
\end{equation}
The integration of this equation along the longitudinal coordinate $z$ renders the trajectories that describe locally how the partially coherent light energy flux redistributes along the transverse direction, in agreement with the effective velocity field $\bar{\bf v}_\perp({\bf r}_\perp,z)$.
These trajectories, based on the CSD, thus generalize the concept of beam trajectory considered in Refs.~\cite{sanz:JOSAA:2022} and \cite{sanz:OE:2024}, since now we have a tool applicable also to partially coherent light beams.
Furthermore, we also note that the expressions for $\bar{I}({\bf r}_\perp,z)$ and $\bar{\bf j}({\bf r}_\perp,z)$ constitute the direct optical analogs, respectively, to the probability density and the quantum flux as obtained from the a quantum system reduced density matrix in the position representation.
Accordingly, the transverse velocity (\ref{eq10}) and the trajectories obtained from it represent the analogous equation of motion and trajectories corresponding to with a (reduced) quantum subsystem \cite{sanz:EPJD:2007}.
This thus makes possible an effective transfer of trajectory-based methodologies and understanding between optical coherence and open quantum systems.


\subsection{Partially coherent Airy beams}
\label{sec22}

The above formulation is general and therefore can be used to describe any type of paraxial partially coherent beam.
As an application, we now study the particular case of one-dimensional Airy beams, without any loss of generality.
To further simplify notation, we are going to consider reduced dimensions in the coordinates, as in Ref.~\cite{sanz:JOSAA:2022}, with the changes $\tilde{x} = x/x_0$ and $\tilde{z} = z/kx_0^2$, where $x_0$ is a typical transverse length value involved in the experiment.
With this change, the amplitude of the Airy beams solutions of the paraxial Helmholtz equation reads \cite{berry:AJP:1979} as
\begin{equation}
\psi(\tilde{x},\tilde{z}) = e^{i(\tilde{x}-\tilde{z}^2/6)\tilde{z}/2}
Ai(\tilde{x}-\tilde{z}^2/4) ,
\end{equation}
with $Ai(s)$ being the usual Airy function \cite{NIST:DLMF}.
%
%
The corresponding CSD reads \cite{sanz:PRA:2022} as
\begin{equation}
W(\tilde{x}',\tilde{x},\tilde{z}) = e^{i(\tilde{x} - \tilde{x}') \tilde{z}/2}
w(\tilde{x}',\tilde{x},\tilde{z}) ,
\label{eq18}
\end{equation}
where
\begin{equation}
w(\tilde{x}',\tilde{x},\tilde{z}) =
\iint \mathcal{C} (\xi, \xi') e^{i(\xi - \xi') \tilde{z}/2}
Ai(\tilde{x}' - \xi' - \tilde{z}^2/4) Ai(\tilde{x} - \xi - \tilde{z}^2/4) d\xi d\xi' .
\label{eq19}
\end{equation}
In Eq.~(\ref{eq19}), the correlation function $\mathcal{C}(\xi, \xi')$ accounts for both the power content (and hence the beam spatial finiteness) and the coherence (see Sec.~\ref{sec30}).
This averaging over realizations of the Airy beam randomly displaced is thus equivalent to the averaging introduced in Eq.~(\ref{eq12}).
The two main questions that we want to answer with this approach is to determine how loss of coherence affects the propagation properties of the beam, and also how coherence and energy finiteness relate in the propagation of the beam.

Following the description in the previous section, the intensity distribution and transverse flux, Eqs.~(\ref{transint}) and~(\ref{eq17}), read as
\begin{eqnarray}
\bar{I}(\tilde{x},\tilde{z}) & = & w(\tilde{x},\tilde{x},\tilde{z}) .
\label{eq21b} \\
\bar{j}(\tilde{x},\tilde{z}) & = &
\frac{\tilde{z}}{2}\ w(\tilde{x},\tilde{x},\tilde{z})
+ \frac{1}{2i} \left[ \frac{\partial w(\tilde{x}',\tilde{x},\tilde{z})}{\partial \tilde{x}}
- \frac{\partial w(\tilde{x}',\tilde{x},\tilde{z})}{\partial \tilde{x}'} \right]_{\tilde{x}'=\tilde{x}} .
\label{eq21a}
\end{eqnarray}
The substitution of Eq.~(\ref{eq19}) into Eqs.~(\ref{eq21a}) and (\ref{eq21b}) renders
\begin{eqnarray}
\bar{I} (\tilde{x},\tilde{z}) & = & \iint \mathcal{C}(\xi, \xi')
\cos [(\xi - \xi') \tilde{z}/2]\ Ai(\tilde{x} - \xi' - \tilde{z}^2/4)
Ai(\tilde{x} - \xi - \tilde{z}^2/4) d\xi d\xi' .
\label{eq24b} \\
\bar{j}(\tilde{x},\tilde{z}) & = &
\frac{\tilde{z}}{2}\ w(\tilde{x},\tilde{x},\tilde{z}) 
- \iint \mathcal{C} (\xi, \xi')\
\sin [(\xi - \xi') \tilde{z}/2]\ Ai(\tilde{x} - \xi' - \tilde{z}^2/4)
\nonumber \\ & & \qquad \qquad \qquad \qquad \qquad \qquad \qquad \qquad \qquad \times
\frac{\partial Ai(\tilde{x} - \xi - \tilde{z}^2/4)}{\partial \tilde{x}}
d\xi' d\xi ,
\label{eq24a}
\end{eqnarray}
Substituting these two results into the effective equation of motion (\ref{eq10}), we obtain
\begin{equation}
	\frac{d\tilde{x}}{d\tilde{z}} = \frac{\tilde{z}}{2}
	- \frac{\displaystyle \iint \mathcal{C}(\xi, \xi') \sin [(\xi - \xi') \tilde{z}/2]\
		Ai(\tilde{x} - \xi' - \tilde{z}^2/4)\
		\frac{\partial Ai(\tilde{x} - \xi - \tilde{z}^2/4)}{\partial \tilde{x}}\
		d\xi' d\xi}
	{\displaystyle \iint \mathcal{C}(\xi, \xi') \cos [(\xi - \xi') \tilde{z}/2]\ Ai(\tilde{x} - \xi' - \tilde{z}^2/4) Ai(\tilde{x} - \xi - \tilde{z}^2/4)
		d\xi' d\xi} .
	\label{eq26}
\end{equation}
This expression constitutes a generalization for partially-coherent finite-energy Airy beams of the one formerly obtained in Refs.~\cite{sanz:JOSAA:2022} and \cite{sanz:OE:2024} for the coherent finite-energy case.
In the high-correlation limit between the random displacements $\xi$ and $\xi'$, i.e., for $C(\xi,\xi') \to \delta(\xi - \xi')$, the second term on the right-hand side of Eq.~(\ref{eq26}) vanishes and the trajectories become parallel parabolas.
This is the behavior that characterizes, at a trajectory level, an ideal Airy beam \cite{sanz:JOSAA:2022}, in compliance with its distinctive properties of transverse shape invariance and self-acceleration along its forward propagation.
Nonetheless, this behavior is much more general and does not depend on whether the beam has full coherence, partial coherence, or is totally incoherent, as long as its intensity distribution still remains characterized by an infinite tail, as it was reported in Ref.~\cite{sanz:PRA:2022}.
It is this long tail, and hence the beam infinite energy content, what keeps a push forward (along the transverse direction) at a constant rate, which in tern gives rise to an overall self-acceleration.
Below we will see that, effectively, even in the case of a lack of coherence, the trajectories will display the typical parabolic behavior provided the intensity distribution tail extends for a long distance.


\subsection{Superposition of two overlapping partially coherent Airy beams}
\label{sec23}

An interesting case worth discussing here is also a situation where we have a partially
coherent superposition of two overlapping Airy beams, with each contribution in
the superposition having a different weight.
As it was shown in Ref.~\cite{sanz:JOSAA:2022}, the overlapping of two Airy beams produces interference
traits, which manifest in a sort of wiggly behavior in the corresponding trajectories.
This thus poses another interesting questions, namely, whether one could expect to recover or observed any typical coherent trait out of a superposition of two poorly coherent beams.
To investigate this fact, consider an initial input superposition of the type
\begin{equation}
\psi(\tilde{x},0) = \cos (\theta/2)\ Ai(\tilde{x}_0 + \tilde{x} - \xi) +
\sin (\theta/2) e^{i\varphi} Ai(\tilde{x}_0 - \tilde{x} + \xi) ,
\label{eq27}
\end{equation}
where we have assumed, for simplicity, the random $\xi$-shift to be the same on both partners.
This superposition has been written in the form of a Bloch vector for a two-state system \cite{nielsen-chuang-bk,breuer-bk:2002,aspect-bk}, where $\theta$ and $\varphi$ denote the polar and azimuthal angles in a sphere (the Bloch sphere, isomorphic to Poincar\'e's sphere) with radius equal to $|\psi|^2$.
Within the present context, the North and South poles of the Bloch sphere represent, respectively, the single propagating Airy beam or the counter-propagating one, while any value around the equator of the sphere describes a qubit-type state, where the two parties contribute exactly the same way to the superposition, although there is still a relative phase between them.

Taking into account the above facts, the propagated beam is of the type
\begin{eqnarray}
\psi(\tilde{x},\tilde{z}) & = & e^{i(\tilde{x}_0 - \tilde{z}^2/6)\tilde{z}/2} \left[ \cos (\theta/2)\ e^{i (\tilde{x} - \xi) \tilde{z}/2}
Ai(\tilde{x}_0 + \tilde{x} - \xi -\tilde{z}^2/4)
 \right. \nonumber \\ &  & \qquad \qquad \qquad \left.
 + \sin (\theta/2) e^{i\varphi} e^{- i (\tilde{x} - \xi) \tilde{z}/2} Ai(\tilde{x}_0 - \tilde{x} + \xi -\tilde{z}^2/4) \right] .
\label{eq27b}
\end{eqnarray}
Proceeding as in the previous section, we obtain the following CSD:
\begin{eqnarray}
W(\tilde{x}',\tilde{x},\tilde{z}) & = &
\iint d\xi d\xi' \mathcal{C} (\xi, \xi') \bigg[
\cos^2 (\theta/2)\ e^{i[(\tilde{x} - \tilde{x}') - (\xi - \xi')] \tilde{z}/2}
\nonumber \\ & & \qquad \qquad \qquad \qquad \qquad \times \phantom{\bigg[}
Ai(\tilde{x}_0 + \tilde{x}' - \xi' - \tilde{z}^2/4)
Ai(\tilde{x}_0 + \tilde{x} - \xi - \tilde{z}^2/4) \nonumber \\
& & + \sin^2 (\theta/2)\ e^{-i[(\tilde{x} - \tilde{x}') - (\xi - \xi')] \tilde{z}/2}
Ai(\tilde{x}_0 - \tilde{x}' + \xi' - \tilde{z}^2/4)
Ai(\tilde{x}_0 - \tilde{x} + \xi - \tilde{z}^2/4) \nonumber \\
& & + \frac{1}{2} \sin \theta e^{-i\varphi} e^{i[(\tilde{x} + \tilde{x}') - (\xi + \xi')] \tilde{z}/2}
Ai(\tilde{x}_0 - \tilde{x}' + \xi' - \tilde{z}^2/4)
Ai(\tilde{x}_0 + \tilde{x} - \xi - \tilde{z}^2/4) \nonumber \\
& & + \frac{1}{2} \sin \theta e^{i\varphi} e^{-i[(\tilde{x} + \tilde{x}') - (\xi + \xi')] \tilde{z}/2}
Ai(\tilde{x}_0 + \tilde{x}' - \xi' - \tilde{z}^2/4)
Ai(\tilde{x}_0 - \tilde{x} + \xi - \tilde{z}^2/4) \bigg] ,
\nonumber \\ & &
\end{eqnarray}
where we can clearly distinguish three contributions, namely, two CSDs each associated with one of
the localized beams (first and second lines) and a crossed interference term (third and forth lines).
Their role is more clearly seen through the corresponding partial contributions, that is,
the transverse intensity distributions,
\begin{eqnarray}
\bar{I}_+(\tilde{x},\tilde{z}) & = &
\cos^2 (\theta/2) \iint d\xi d\xi' \mathcal{C} (\xi, \xi')
\cos [(\xi - \xi') \tilde{z}/2]\
\nonumber \\ & & \qquad \qquad \qquad \qquad \times
Ai(\tilde{x}_0 + \tilde{x} - \xi' - \tilde{z}^2/4)
Ai(\tilde{x}_0 + \tilde{x} - \xi - \tilde{z}^2/4) , \\
\bar{I}_-(\tilde{x},\tilde{z}) & = &
\sin^2 (\theta/2) \iint d\xi d\xi' \mathcal{C} (\xi, \xi')
\cos [(\xi - \xi') \tilde{z}/2]\
\nonumber \\ & & \qquad \qquad \qquad \qquad \times
Ai(\tilde{x}_0 - \tilde{x} + \xi' - \tilde{z}^2/4)
Ai(\tilde{x}_0 - \tilde{x} + \xi - \tilde{z}^2/4) , \\
\bar{I}_{\rm int} (\tilde{x},\tilde{z}) & = &
\sin \theta \iint d\xi d\xi' \mathcal{C} (\xi, \xi')
\cos [\tilde{x} \tilde{z} - (\xi + \xi') \tilde{z}/2 - \varphi]\
\nonumber \\ & & \qquad \qquad \qquad \qquad \times 
Ai(\tilde{x}_0 + \tilde{x} - \xi' - \tilde{z}^2/4)
Ai(\tilde{x}_0 - \tilde{x} + \xi - \tilde{z}^2/4) ,
\end{eqnarray}
and the respective transverse fluxes,
\begin{eqnarray}
\bar{j}_+(\tilde{x},\tilde{z}) & = &
\cos^2 (\theta/2) \left\{ \frac{\tilde{z}}{2}\ \langle |\psi_+(\tilde{x},\tilde{z})|^2 \rangle
 \right. \nonumber \\ & & \left. \qquad \qquad \quad
- \iint d\xi d\xi' \mathcal{C} (\xi, \xi') \sin [(\xi - \xi') \tilde{z}/2]\
 \right. \nonumber \\ & & \left. \qquad \qquad \qquad \times
Ai(\tilde{x}_0 + \tilde{x} - \xi' - \tilde{z}^2/4)
\frac{\partial Ai(\tilde{x}_0 + \tilde{x} - \xi - \tilde{z}^2/4)}{\partial \tilde{x}} \right\} , \\
\bar{j}_-(\tilde{x},\tilde{z}) & = &
\sin^2 (\theta/2) \left\{ - \frac{\tilde{z}}{2}\ \langle |\psi_-(\tilde{x},\tilde{z})|^2 \rangle
 \right. \nonumber \\ & & \left. \qquad \qquad \quad
+ \iint d\xi d\xi' \mathcal{C} (\xi, \xi') \sin [(\xi - \xi') \tilde{z}/2]\
 \right. \nonumber \\ & & \left. \qquad \qquad \qquad \times
Ai(\tilde{x}_0 - \tilde{x} + \xi' - \tilde{z}^2/4)
\frac{\partial Ai(\tilde{x}_0 - \tilde{x} + \xi - \tilde{z}^2/4)}{\partial \tilde{x}} \right\} ,
\end{eqnarray}

\begin{eqnarray}
\bar{j}_{\rm int} (\tilde{x},\tilde{z}) & = &
\frac{1}{2}\ \sin \theta \iint d\xi d\xi' \mathcal{C} (\xi, \xi')
\sin [\tilde{x} \tilde{z} - (\xi + \xi') \tilde{z}/2 - \varphi]\
\nonumber \\ & &  \qquad \quad \times
\left\{  Ai(\tilde{x}_0 - \tilde{x} + \xi' - \tilde{z}^2/4)
\frac{\partial Ai(\tilde{x}_0 + \tilde{x} - \xi - \tilde{z}^2/4)}{\partial \tilde{x}}
 \right. \nonumber \\ & & \left. \qquad \qquad \qquad
- \frac{\partial Ai(\tilde{x}_0 - \tilde{x} + \xi' - \tilde{z}^2/4)}{\partial \tilde{x}}
Ai(\tilde{x}_0 + \tilde{x} - \xi - \tilde{z}^2/4) \right\} .
\end{eqnarray}
Note that the expression for the transverse velocity now is not going to be as simple as in the previous cases, since it consists of three intertwined contributions
\begin{equation}
\frac{d\tilde{x}}{d\tilde{z}} = \frac{\bar{j}_+(\tilde{x},\tilde{z})
	+ \bar{j}_-(\tilde{x},\tilde{z}) + \bar{j}_{\rm int}(\tilde{x},\tilde{z})}
{\bar{I}_+(\tilde{x},\tilde{z}) + \bar{I}_-(\tilde{x},\tilde{z})+ \bar{I}_{\rm int} (\tilde{x},\tilde{z})} .
\label{eqmotint}
\end{equation}

Despite of the complexity displayed by the above equation of motion, some particular
properties or behaviors can still be inferred from it.
For instance, it is clear that, while at the poles of the Bloch sphere the beam describes
a single contribution, either propagating forward (along positive $x$, for $\theta = 0$)
or backwards ($\theta = \pi$), the equator of the sphere ($\theta = \pi/2$, for all
$\varphi$) represents all possible situations where we have an equal (unbiased)
contribution from both overlapping beams.
In this latter case, we note that, if the peak-to-peak distance between the main maxima of
the two wave packets is relatively large, at relatively small values of $z$, the
trajectories will correspond essentially to those of separated Airy beams, since
Eq.~(\ref{eqmotint}) can be approximated by
\begin{equation}
\frac{d\tilde{x}_\pm}{d\tilde{z}} \approx
\frac{\bar{j}_\pm(\tilde{x},\tilde{z})}{\bar{I}_\pm(\tilde{x},\tilde{z})} ,
\label{eqmotint2}
\end{equation}
where the $\pm$ sign denotes that the trajectories belong either to the plus or the minus
contribution.

On the other hand, when both beams strongly overlap, the interference traits observed are
going to depend importantly on the crossed or interference term (denoted by ``int''), which
appears both in the numerator and in the denominator, and hence $\varphi$ plays a role.
That is, even in the case of dealing with partially coherent beams, interference-like
traits are expected for $z$ distances beyond the point where both contributions overlap.
In this case, if $\varphi=0$, there will be a maximum at the center ($x=0$), while for
$\varphi = \pi$, the center of the intensity distribution will be a minimum.
In this latter case, the trajectories will avoid approaching $x=0$, while in the former
case they will converge to it, as shown in Ref.~\cite{sanz:JOSAA:2022} for the fully coherent
case.
Now, if $\varphi=\pi/2$ or $\varphi=3\pi/2$, there will not be either a maximum or a minimum,
but just a $\pi/2$ shift towards the right or the left of $x=0$, respectively.


\section{Results}
\label{sec3}

\subsection{Correlation function}
\label{sec30}

In analogy to the cases investigated in Ref.~\cite{sanz:PRA:2022}, here we are also going to consider a situation with quasi-infinite energy but poor coherence, as well as the opposite situation, that is, finite energy but good coherence.
Specifically, we consider a $\mathcal{C} (\xi, \xi')$ function with the functional form
\begin{equation}
	\mathcal{C} (\xi, \xi') = \frac{\sqrt{\mu^2 + 4\sigma^2}}{2\pi\sigma\mu} e^{-(\xi^2 + \xi'^2)/2\sigma^2} e^{-(\xi - \xi')^2/\mu^2} ,
	\label{eq20b}
\end{equation}
such that $\int \mathcal{C} (\xi, \xi') d\xi d\xi' = 1$.
This correlation function is similar to the degree of coherence introduced by Starikov and Wolf to study the mode structure of a Gaussian Schell-model source \cite{schell1961-phd,gori:OptCommun:1980,wolf:OPL:1978,wolf:JOSA:1982,friberg:OptCommun:1982}.
Although both the energy content and the global coherence depend on both $\sigma$ and $\mu$, for a qualitative but highly illustrative analysis we make a series of considerations.
Note that, if $\mu \to 0$, then the second term approaches Dirac's delta function, and $\mathcal{C} (\xi, \xi')$ reduces to
\begin{equation}
	\mathcal{C} (\xi) \to \frac{1}{\sqrt{\pi\sigma^2}} e^{-\xi^2/2\sigma^2} ,
\end{equation}
which is the $P(\xi)$ Gaussian spread function considered in Ref.~\cite{sanz:PRA:2022}.
Although in this case the energy content (in terms of the beam tail extension) is still infinite, depending on the value of $\sigma$ the beam coherence will be more or less affected \cite{wolf:JOSA:1982}.
In particular, for small $\sigma$ the dispersion of $\xi$ values will also be small, and hence the beam global coherence will be preserved.
On the other hand, for large $\sigma$, the dispersion of $\xi$ values will be larger, which implies an important loss of coherence (or even its total suppression) and manifests as a loss of the distinctive oscillatory behavior of the beam tail.
Finite values of $\mu$ imply finite-energy conditions, which manifests as a decrease in the beam tail extent.
In such cases, depending on the whether $\sigma$ is small or large, we may have good or poor coherence, that is, the oscillatory behavior of the tail will present a better visibility or not, respectively.
Below, we analyze the behaviors associated with the following sets of values, which are representative of the three main scenarios that we can find:
\begin{itemize}
	\item {\bf Case 1:} $\sigma = 1$ and $\mu = 0.125 \sqrt{2}$, which describes a situation with a large spatial extension (quasi-infinite energy), but very poor coherence.
	
	\item {\bf Case 2:} $\sigma = 0.5$ and $\mu = 2 \sqrt{2}$, which describes a situation with a shorter spatial extension (finite energy), but good coherence.
	
	\item {\bf Case 3:} $\sigma = 1$ and $\mu = 2 \sqrt{2}$, which describes an intermediate situation.
\end{itemize}


\subsection{Numerical details}

Apart from the above considerations on the correlation function $\mathcal{C}(\xi,\xi')$, in the results presented here we have made used of own on-purpose prepared beam propagation codes relying on the split-operator method, formerly proposed by Feit and Fleck \cite{feit-fleck:ApplOpt:1978,feit-fleck:ApplOpt:1980-1,feit-fleck:ApplOpt:1980-2,feit-fleck:ApplOpt:1980-3,leforestier:jcompphys:1991}, combined with the efficiency of the fast Fourier algorithm to both solving the spatial derivatives \cite{feit-fleck:JCompPhys:1982,kosloff:JCP:1983,kosloff:JCompPhys:1983} and, at the same time, computing the energy flow trajectories \cite{sanz-bk-2,sanz:JOSAA:2022}.
In all calculations, we have monitored the transverse flux by means of sets of 31 trajectories with evenly spaced initial conditions between $\tilde{x} = -10$ and $\tilde{x} = 3.5$ (in reduced units), although neither all regions of the initial distribution are equally dense populated (the energy distributes unevenly) nor the forefront trajectories are associated with a region particularly intense (the intensity distribution beyond $\tilde{x} \approx 1.5$ is very low).
This choice allows us to get an overall map of the propagation dynamics in the three cases described above.
The same holds for the results associated with the overlapping partially coherent Airy beams, where twice the number of trajectories has been considered, with symmetric initial positions with respect to $x=0$.


\begin{figure}[!t]
	\centering
	\includegraphics[width=\columnwidth]{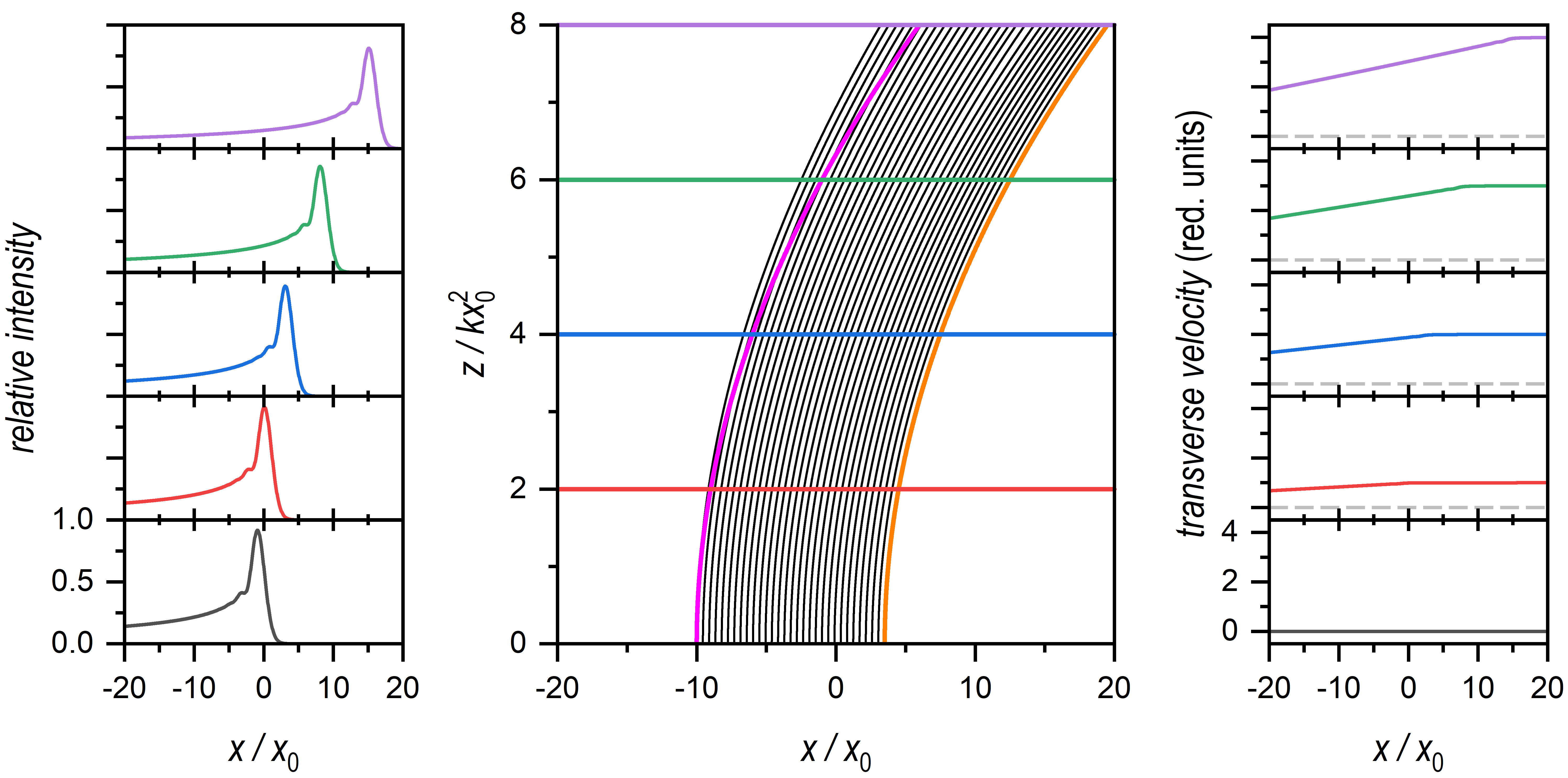}
	\caption{\label{fig1}
		Propagation of a partially coherent beam with a CSD characterized by $\sigma = 1$ and $\mu = 0.125 \sqrt{2}$ (case~1; see text for details) up to $z$ near 30~cm.
		{\bf Left:} intensity distributions at different values of $z$.
		In reduced units, $\tilde{z} = z/kx_0^2$: $\tilde{z}=0$ (black), $\tilde{z}=2$ (red),
		$\tilde{z}=4$ (blue), $\tilde{z}=6$ (green), and $\tilde{z}=8$ (purple).
		{\bf Center:} Set of 31 trajectories with evenly spaced initial positions between $\tilde{x}=-10$
		and $\tilde{x}=3.5$ (covering a range of about 0.7~mm).
		The magenta and orange lines represent the trajectories for an ideal Airy beam with initial
		positions at the margins of the trajectory distribution.
		Vertical colored lines correspond to the $z$ values for the intensity distributions and transverse
		velocity fields displayed on the left and the right panels.
		{\bf Right:} transverse velocity field calculated at the same $z$ values as the intensity
		distribution profiles displayed on the left panels.
		The zero-velocity condition is indicated with the gray dashed line.
	}
\end{figure}

\subsection{Propagation of partially coherent Airy beams}
\label{sec31}

Case~1 is represented in Fig.~\ref{fig1}.
The trajectories are shown in the central panel, while the two marginal column panels
show the intensity distribution (left) and the transverse velocity (right) at different
values of $\tilde{z}$ ($\tilde{z} = 0, 2, 4, 6, 8$), which cover the range between the
input plane and $z \approx 300$~mm.
The color of the horizontal lines in the trajectory plot are in correspondence with the
plots displayed in the column panels for an easier identification.
Furthermore, the trajectories corresponding to an ideal Airy beam are represented with
the magenta and orange lines in the trajectory plot to better understand how much the
trajectories for the partially coherent beam deviate from those corresponding to an ideal
one \cite{sanz:JOSAA:2022}.
As it can be noticed, even a remarkable lack of coherence does not prevent the partially
coherent beam from keeping its properties of shape invariance and self-acceleration along
the transverse direction for a rather long propagation distance.
The snapshots at different $z$-values of the  intensity distribution, on the left, show how
these properties are preserve.
However, more importantly, the fact that the selected swarm of trajectories essentially
remains confined within the trajectories for the ideal beam clearly shows that there is
a trend in these beams to keep such properties for long distances.
Actually, this property is better preserved in the front part than in the rear part, where
trajectories show a lower acceleration, thus gradually getting delayed with respect to the
trajectory for the ideal Airy beam (magenta line).
If we inspect the profiles of the transverse velocity at the same $z$-values, on the right,
we readily find the explanation for this behavior: the incipient appearance of a slope in
the rear part of the beam, which increases as $z$ also increases.
This decrease implies lower values of the velocity and hence a slower acceleration for the
trajectories in the rear part of the swarm.

\begin{figure}[!t]
	\centering
	\includegraphics[width=\columnwidth]{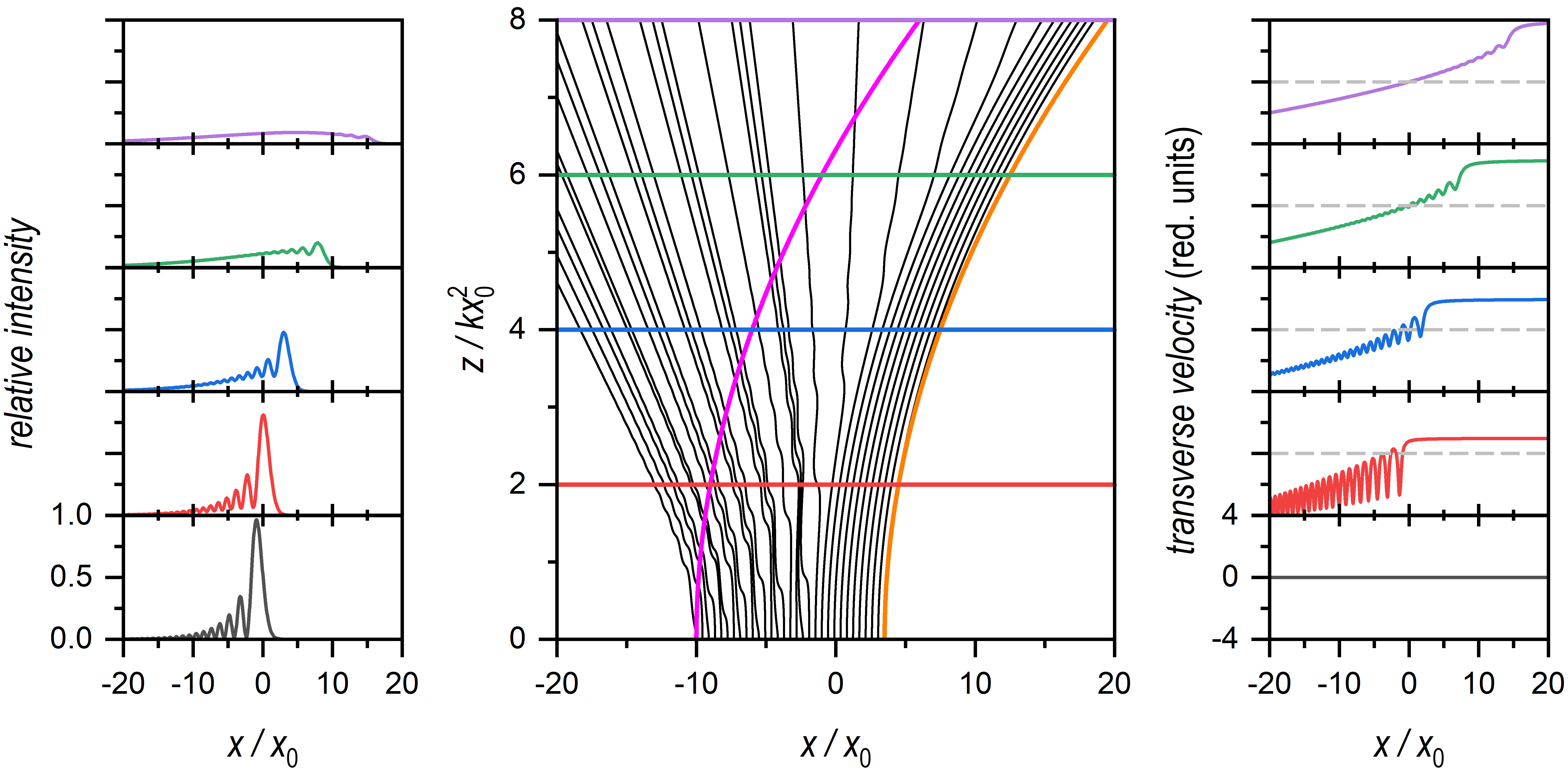}
	\caption{\label{fig2}
		Propagation of a partially coherent beam with a CSD characterized by $\sigma = 0.5$ and $\mu = 2 \sqrt{2}$ (case~2; see text for details) up to $z$ near 30~cm.
		{\bf Left:} intensity distributions at different values of $z$.
		In reduced units, $\tilde{z} = z/kx_0^2$: $\tilde{z}=0$ (black), $\tilde{z}=2$ (red),
		$\tilde{z}=4$ (blue), $\tilde{z}=6$ (green), and $\tilde{z}=8$ (purple).
		{\bf Center:} Set of 31 trajectories with evenly spaced initial positions between $\tilde{x}=-10$
		and $\tilde{x}=3.5$ (covering a range of about 0.7~mm).
		The magenta and orange lines represent the trajectories for an ideal Airy beam with initial
		positions at the margins of the trajectory distribution.
		Vertical colored lines correspond to the $z$ values for the intensity distributions and transverse
		velocity fields displayed on the left and the right panels.
		{\bf Right:} transverse velocity field calculated at the same $z$ values as the intensity
		distribution profiles displayed on the left panels.
		The zero-velocity condition is indicated with the gray dashed line.
	}
\end{figure}

If we now analyze what happens in a case with finite energy, but good coherence, as it is
the situation with case~2, the corresponding profiles of the transverse velocity show a
different trend, as seen in Fig.~\ref{fig2}.
In this case, the snapshots of the intensity distribution show an apparently fast decay of
the Airy beam properties beyond $z \approx 150$~mm.
The trajectories, though, allow us to observe that, although the swarm tends to distribute
in a sort of Gaussian way, suppressing any information about presence of maxima or minima,
there is still a subgroup of trajectories that clearly exhibits self-acceleration (those
closer to the orange trajectory of the ideal Airy beam).
To understand this behavior, note in the profiles of the transverse velocity (see right
panels in Fig.~\ref{fig2}) the existence of two different regimes for nonzero $z$, denoted
by a plateau at the front and a decaying tail in the rear part.
The constant value of the velocity at the plateau allows the trajectories that it affects
to display similar features, which translate in a nearly parallel forward motion.
On the other hand, the decaying tail will provoke that the trajectories affected will
display a very different behavior.
At short $z$ values, the oscillatory tail will give rise to the appearance of a wiggling
motion; at long $z$ values, where the tail is basically monotonically decreasing, wiggles
will disappear and the motion will be similar to the one observed in case~1.

\begin{figure}[!t]
	\centering
	\includegraphics[width=\columnwidth]{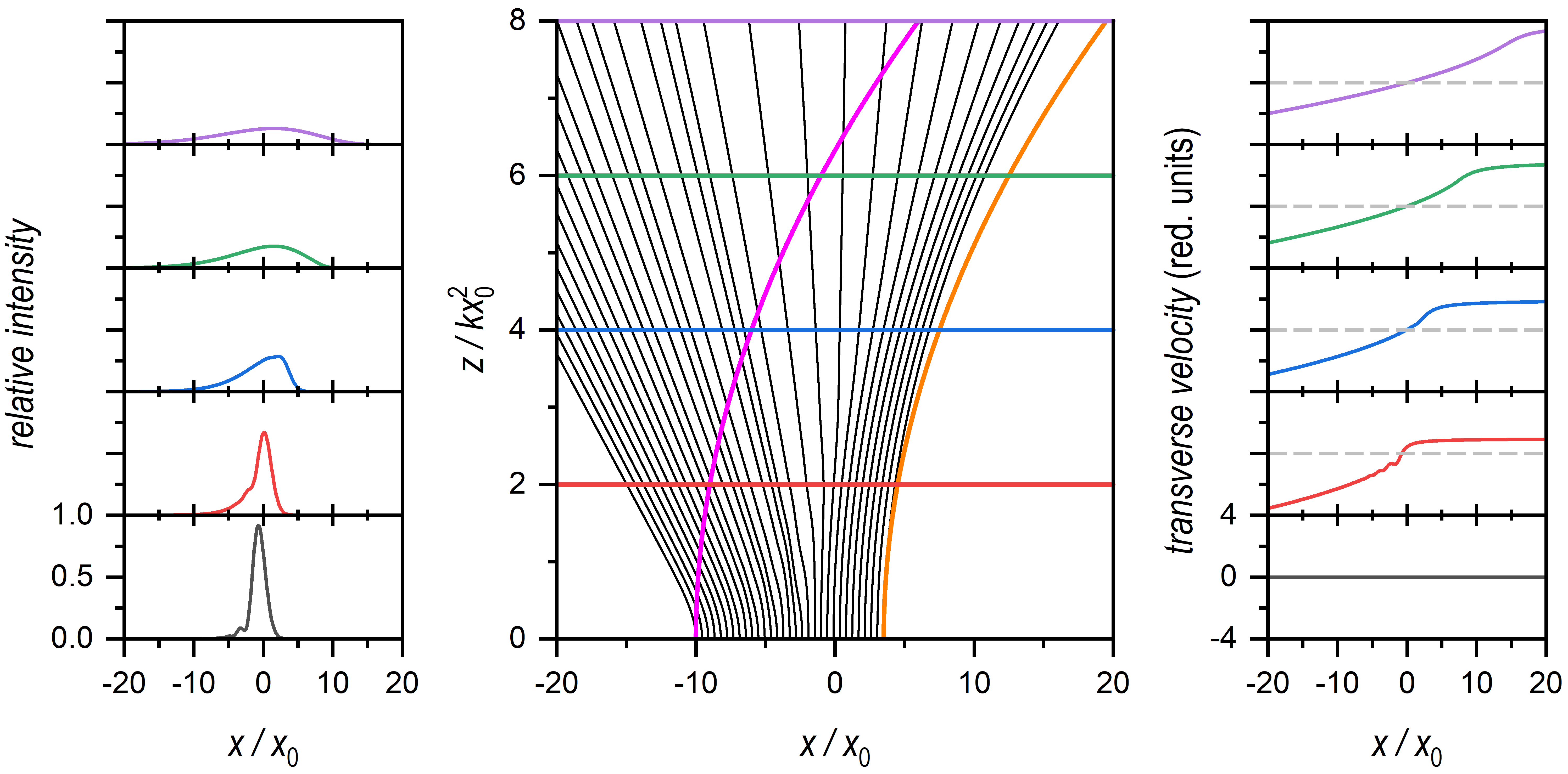}
	\caption{\label{fig3}
		Propagation of a partially coherent beam with a CSD characterized by $\sigma = 1$ and $\mu = 2 \sqrt{2}$ (case~3; see for text for details) up to $z$ near 30~cm.
		{\bf Left:} intensity distributions at different values of $z$.
		In reduced units, $\tilde{z} = z/kx_0^2$: $\tilde{z}=0$ (black), $\tilde{z}=2$ (red),
		$\tilde{z}=4$ (blue), $\tilde{z}=6$ (green), and $\tilde{z}=8$ (purple).
		{\bf Center:} Set of 31 trajectories with evenly spaced initial positions between $\tilde{x}=-10$
		and $\tilde{x}=3.5$ (covering a range of about 0.7~mm).
		The magenta and orange lines represent the trajectories for an ideal Airy beam with initial
		positions at the margins of the trajectory distribution.
		Vertical colored lines correspond to the $z$ values for the intensity distributions and transverse
		velocity fields displayed on the left and the right panels.
		{\bf Right:} transverse velocity field calculated at the same $z$ values as the intensity
		distribution profiles displayed on the left panels.
		The zero-velocity condition is indicated with the gray dashed line.
	}
\end{figure}

A situation in between the above two cases is that represented by case~3.
This situation is represented in Fig.~\ref{fig3}, where we can see that not only the
shape invariance, but also the self-accelerated motion are suppressed very quickly.
The intensity distribution approaches a kind of asymmetric Gaussian distribution,
while the trajectories show a nearly homogeneous separation, which somehow mimics
the behavior exhibited by trajectories associated with a Gaussian beam.
Actually, if we observe the transverse velocity profiles, on the right panels, we find
a trend to approach a nearly linear variation with the transverse coordinate, which is
precisely the transverse field corresponding to a Gaussian beam with zero transverse
momentum (velocity).

\begin{figure}[!b]
	\centering
	\includegraphics[width=\columnwidth]{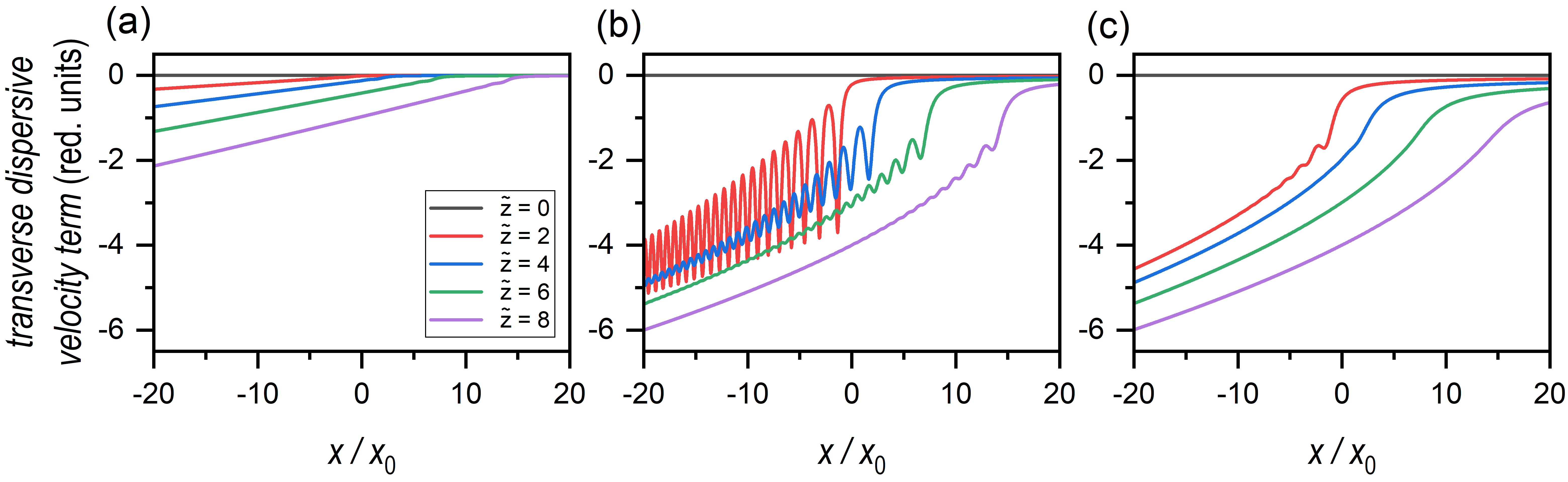}
	\caption{\label{fig4}
		Dispersive part of Eq.~(\ref{eq26}) for partially coherent beams with a CSD characterized by:
		(a) $\sigma = 1$ and $\mu = 0.125 \sqrt{2}$ (case~1), (b) $\sigma = 0.5$ and $\mu = 2 \sqrt{2}$ (case~2), and (c) $\sigma = 1$ and $\mu = 2 \sqrt{2}$ (case~3).
		In reduced units, $\tilde{z} = z/kx_0^2$: $\tilde{z}=0$ (black), $\tilde{z}=2$ (red),
		$\tilde{z}=4$ (blue), $\tilde{z}=6$ (green), and $\tilde{z}=8$ (purple). 
	}
\end{figure}

As we can see in Eq.~(\ref{eq26}), the transverse velocity field consists of two well-defined contributions, one which is related to the beam (transverse) propagation, depending linearly on $z$, and another related to its dispersion, which contains information of both the finiteness of the energy and the beam coherence.
Hence, in order to better understand the contribution from these two latter aspects, in
Fig.~\ref{fig4} we have plotted only the dispersive part of the transverse velocity for
the three cases here considered at the corresponding $z$-values.
Note that by “dispersive part” of the transverse velocity we refer to the second term on the right-hand side of Eq.~(\ref{eq26}), which is the one responsible for shortening the back tail and, therefore, causing the gradual loss of ideal Airy beam properties as $z$ increases.
Thus, the first remarkable feature that we find is that, in all cases, the plateau always coincides with a zero-velocity regime.
Accordingly, all trajectories seating on this region will only be propelled by the propagating part of Eq.~(\ref{eq26}), thus displaying the well-known self-accelerated motion.
As for the decaying tale, note that it goes gradually into more negative values of the
velocity, which will counterbalance the forward acceleration imprinted by the linear term.
In cases~1 and~3, this trend is monotonic, although the cancellation in case~3 is going to be faster.
In case~2, however, although the cancellation is also relatively fast, the same trend as in the two previous cases will only be asymptotically, for large $z$, due to the presence of oscillations for short $z$.


\subsection{Interference and partially coherent Airy beams}
\label{sec32}

In order to evaluate now how the finiteness of the energy or a decrease in the coherence affects
a CSD consisting of a partially coherent superposition of two overlapping Airy beams, as described in Theory, let us consider the same three cases as before, but assuming that the peak-to-peak distance between the main maxima of those two contributions is of about 0.6~mm ($\sim 11.3$ in reduced units).
To this end, we consider $x_0 = 0.246$~mm in Eq.~(\ref{eq27}).
Moreover, each contribution of the CSD will be mapped by a total of 31 trajectories, also evenly
distributed between $-10 - x_0$ and $3.5 - x_0$, and $-3.5 + x_0$ and $10 + x_0$, respectively.
For simplicity in the analysis here, we are going to focus on superpositions built on the point
on the Bloch sphere $\theta = \pi/2$ and $\varphi = 0$, given that any other situation either
gives us a biased superposition $\theta \ne \pi/2$ or produces a shift of the interference traits
($\varphi \ne 0$, due to the dependence of the crossed interference terms on $\varphi$).

\begin{figure}[!t]
	\centering
	\includegraphics[width=\columnwidth]{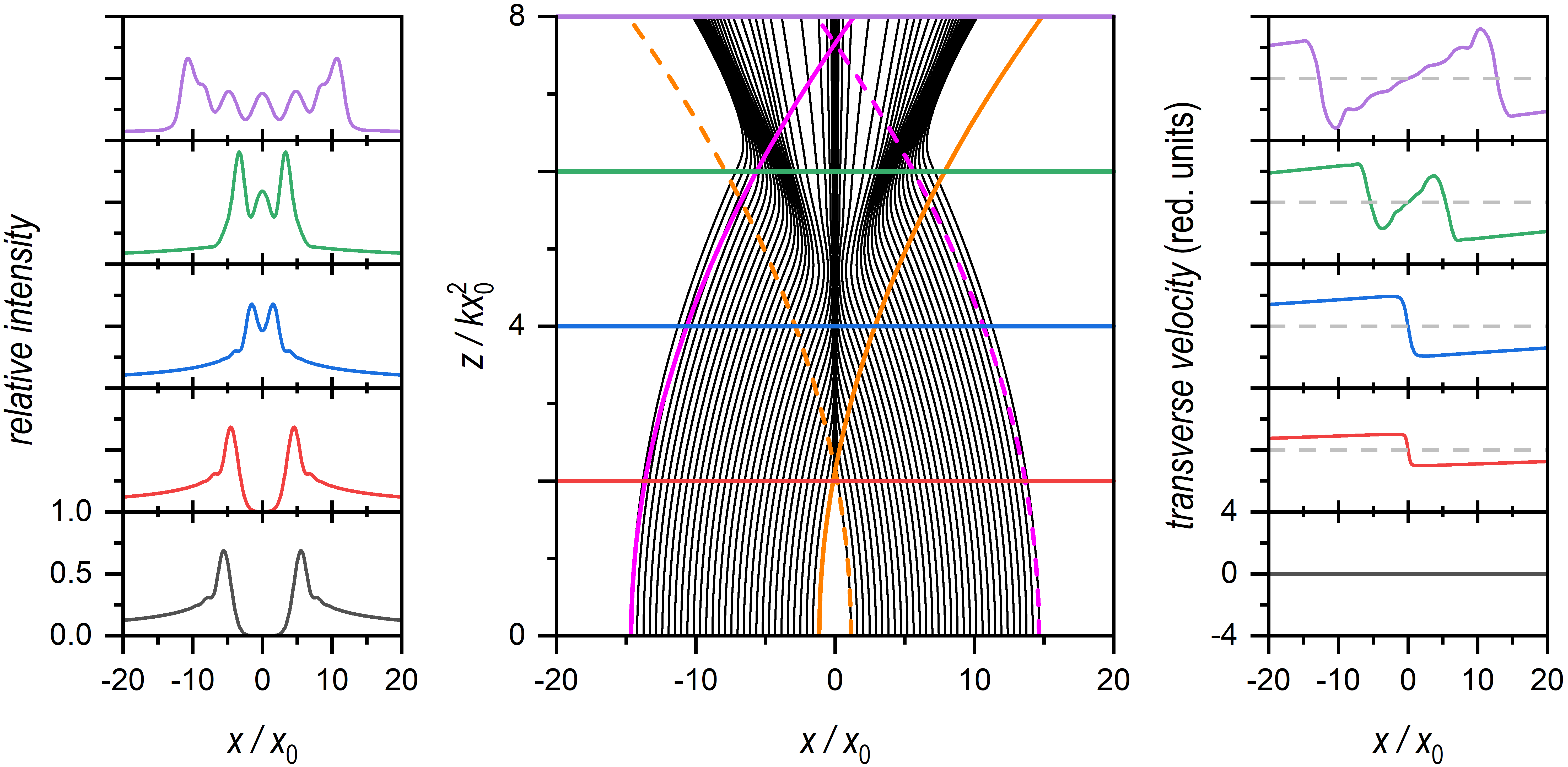}
	\caption{\label{fig5}
		Propagation of a partially coherent beam with a CSD consisting of two contributions, and
		characterized by $\sigma = 1$ and $\mu = 0.125 \sqrt{2}$ (case~1; see text for details) up to $z$ near 30~cm.
		{\bf Left:} intensity distributions at different values of $z$.
		In reduced units, $\tilde{z} = z/kx_0^2$: $\tilde{z}=0$ (black), $\tilde{z}=2$ (red),
		$\tilde{z}=4$ (blue), $\tilde{z}=6$ (green), and $\tilde{z}=8$ (purple).
		{\bf Center:} Set of 31 trajectories with evenly spaced initial positions between $\tilde{x}=-10$
		and $\tilde{x}=3.5$ (covering a range of about 0.7~mm).
		The magenta and orange lines represent the trajectories for an ideal Airy beam with initial
		positions at the margins of the trajectory distribution (with solid line for the contribution with
		positive acceleration and with dashed line for the one with negative acceleration).
		Vertical colored lines correspond to the $z$ values for the intensity distributions and transverse
		velocity fields displayed on the left and the right panels.
		{\bf Right:} transverse velocity field calculated at the same $z$ values as the intensity
		distribution profiles displayed on the left panels.
		The zero-velocity condition is indicated with the gray dashed line.
	}
\end{figure}

In Fig.~\ref{fig5}, for case~1, we observe that, in spite of have a lack of oscillatory structure
before the two contributions merge, afterwards we note the appearance of interference-like traits
in the intensity distribution, within the boundaries determined by the leading maxima moving in
opposite directions.
These oscillations are associated with an uneven distribution of the trajectories after the two
swarms have reached their maximum approach (see waist of the ensemble form by the two swarms at
$\tilde{z} \approx 6$).
This behavior can be better understood by inspecting the transverse velocity profiles, on the right
panels.
Before the coalescence of the two contributions, there is a clear (sudden) jump between a region with
positive momentum (on the left) and another with negative momentum (on the right), which make those
two contributions to move against each other.
After the full overlapping of the two leading maxima, the profile of the transverse velocity undergoes
an important change: the regions with positive and negative momentum start getting apart, and central
region with positive slope emerges in between.
This new region, with nearly evenly distributed oscillations, is responsible for the appearance of the
interference-like traits.

\begin{figure}[!t]
	\centering
	\includegraphics[width=\columnwidth]{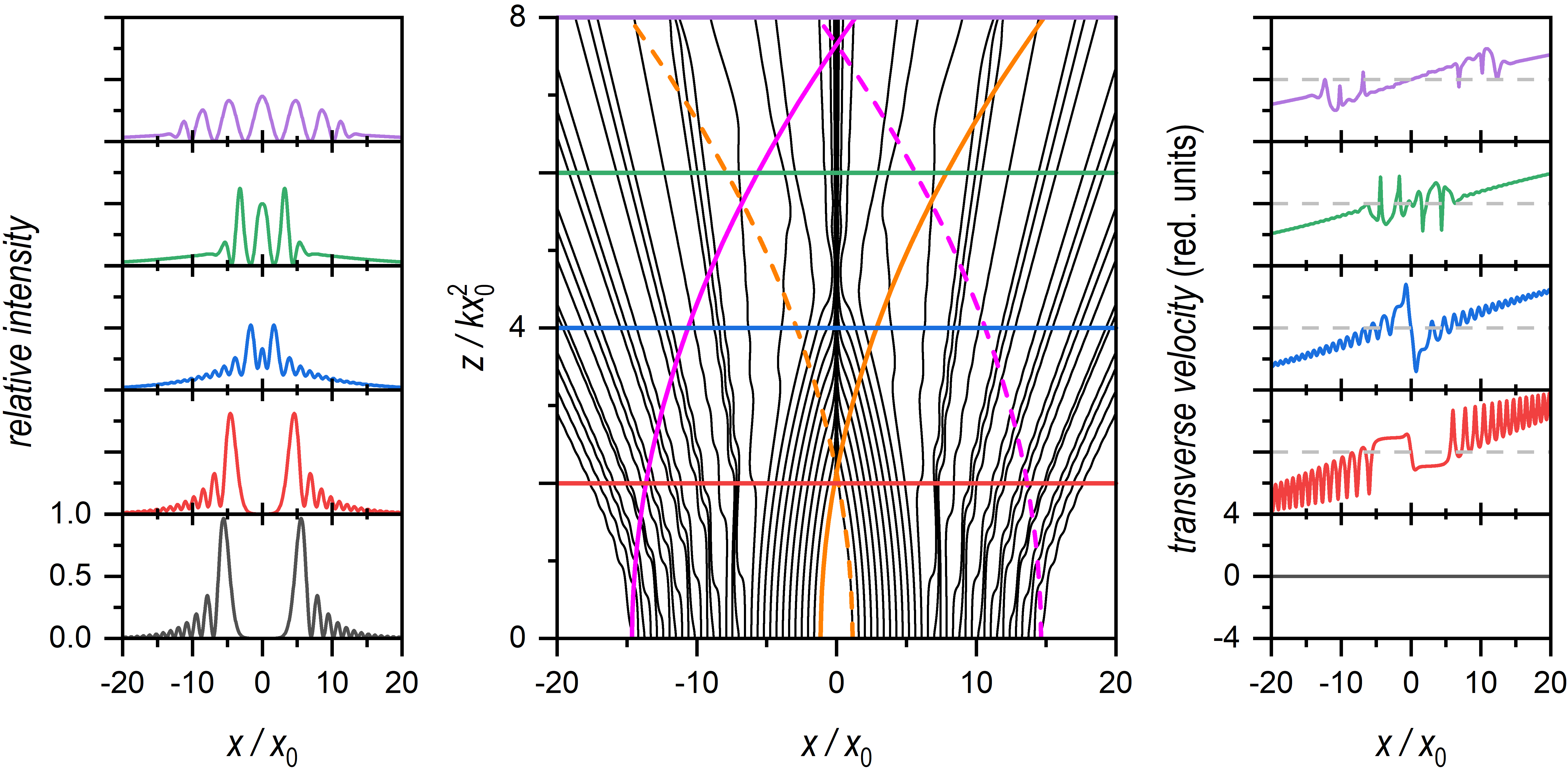}
	\caption{\label{fig6}
		Propagation of a partially coherent beam with a CSD consisting of two contributions, and
		characterized by $\sigma = 0.5$ and $\mu = 2 \sqrt{2}$ (case~2; see text for details) up to $z$ near 30~cm.
		{\bf Left:} intensity distributions at different values of $z$.
		In reduced units, $\tilde{z} = z/kx_0^2$: $\tilde{z}=0$ (black), $\tilde{z}=2$ (red),
		$\tilde{z}=4$ (blue), $\tilde{z}=6$ (green), and $\tilde{z}=8$ (purple).
		{\bf Center:} Set of 31 trajectories with evenly spaced initial positions between $\tilde{x}=-10$
		and $\tilde{x}=3.5$ (covering a range of about 0.7~mm).
		The magenta and orange lines represent the trajectories for an ideal Airy beam with initial
		positions at the margins of the trajectory distribution (with solid line for the contribution with
		positive acceleration and with dashed line for the one with negative acceleration).
		Vertical colored lines correspond to the $z$ values for the intensity distributions and transverse
		velocity fields displayed on the left and the right panels.
		{\bf Right:} transverse velocity field calculated at the same $z$ values as the intensity
		distribution profiles displayed on the left panels.
		The zero-velocity condition is indicated with the gray dashed line.
	}
\end{figure}

\begin{figure}[!t]
	\centering
	\includegraphics[width=\columnwidth]{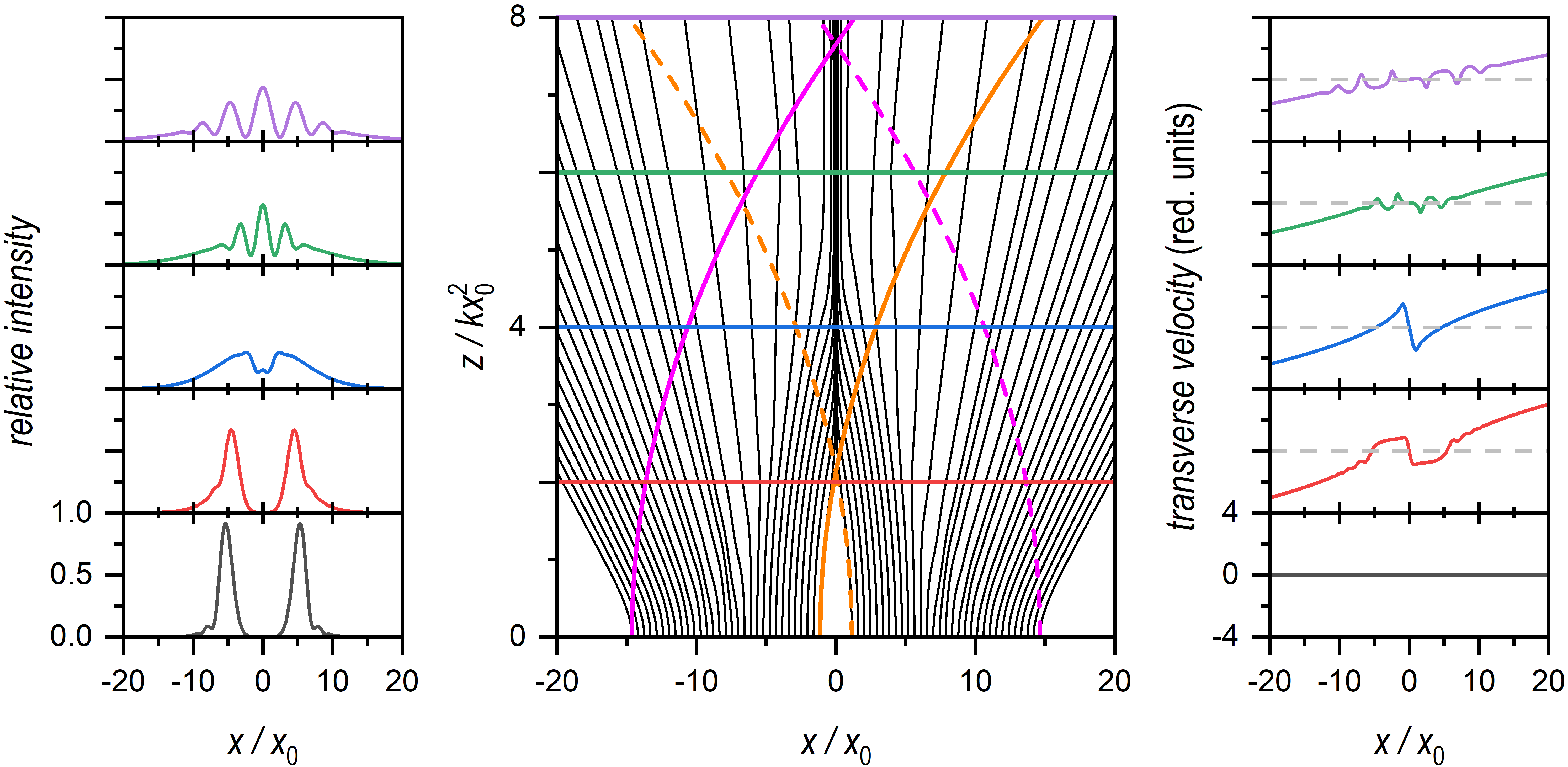}
	\caption{\label{fig7}
		Propagation of a partially coherent beam with a CSD consisting of two contributions, and
		characterized by $\sigma = 1$ and $\mu = 2 \sqrt{2}$ (case~3; see text for details) up to $z$ near 30~cm.
		{\bf Left:} intensity distributions at different values of $z$.
		In reduced units, $\tilde{z} = z/kx_0^2$: $\tilde{z}=0$ (black), $\tilde{z}=2$ (red),
		$\tilde{z}=4$ (blue), $\tilde{z}=6$ (green), and $\tilde{z}=8$ (purple).
		{\bf Center:} Set of 31 trajectories with evenly spaced initial positions between $\tilde{x}=-10$
		and $\tilde{x}=3.5$ (covering a range of about 0.7~mm).
		The magenta and orange lines represent the trajectories for an ideal Airy beam with initial
		positions at the margins of the trajectory distribution (with solid line for the contribution with
		positive acceleration and with dashed line for the one with negative acceleration).
		Vertical colored lines correspond to the $z$ values for the intensity distributions and transverse
		velocity fields displayed on the left and the right panels.
		{\bf Right:} transverse velocity field calculated at the same $z$ values as the intensity
		distribution profiles displayed on the left panels.
		The zero-velocity condition is indicated with the gray dashed line.
	}
\end{figure}

The interference features will be more pronounced depending on the oscillations observed in the
inter main-maxima region.
This can be better appreciated in the results for cases~2 and~3, represented in Figs.~\ref{fig6}
and~\ref{fig7}, respectively, where, after overlapping, we note the presence of a series of
interference maxima with an excellent visibility (see intensity distributions for large $z$ in
the left panels in both figures).

\section{Final remarks}
\label{sec4}

Here we have presented a robust trajectory-flow formulation for paraxial partially coherent beams, grounded on the cross-spectral density, which includes, as a particular case, the description of fully coherent beams.
Specifically, the two key elements are the effective transverse velocity field (acting as a global descriptor) and the flow trajectories emanating from the latter (which provide us with a local, event-by-event description of the propagation process).
These two elements thus supplement the information provided by the transverse intensity distribution.
Physically, this approach stresses the role of the spatial phase variations along the transverse direction, which rule the beam propagation and, therefore, enable a better understanding of the specific features involved and/or developed along such propagation.
It is worth noting that this perspective on the propagation of partially coherent beams constitutes a direct optical analog of the description in terms of the reduced density matrix for nonzero mass and spinless open quantum systems. 

Although the formalism here developed is of general validity, we have applied it to investigate the propagation of a specific type of partially coherent Airy beams.
In particular, the trajectories show that, even under conditions of poor coherence, as long as a partially coherent Airy beam tail remains long enough (i.e., the beam remains quasi-infinite in its energy content), it will still display the renowned properties of ideal Airy beams along the transverse direction during their propagation: shape invariance and quadratic displacement (self-acceleration).
On the contrary, if the beam has a finite energy content, even if it is highly coherent, the properties of ideal Airy beam are lost relatively quickly, depending on how much the beam has been truncated.
In such a case, the trajectories help us to determine quantitatively for how long such properties are still preserved.
In this regard, we find that the transverse velocity field is very useful to quantitatively determine when the beam is going to lose its ideal properties, since this field presents two clearly distinctive behaviors in the case of an ideal or quasi-ideal beam: the rear part is characterized by a decaying slope (with an oscillatory behavior for coherent beams), while the forefront one corresponds to a plateau that moves forward linearly with the propagation coordinate.

In order to further investigate the persistence of coherence traits even in the poorest coherence conditions, we have also considered the overlapping of partially coherent Airy beams with the features of the three instances here analyzed.
It can be observed that, regardless of the coherence content, in all cases an interference-type light distribution arises beyond the point where the two leading maxima maximally overlap.
The energy content, though, introduces a major difference between the light distribution that will be obtained.
If the energy content is quasi-infinite, regardless of the coherence of the initial superposition, the structure of alternating maxima and minima in the light distribution will be confined between the positions of the two leading maxima.
On the contrary, if the energy content is finite, such that the back tail of the beams is rather limited in extension, what we will observe is a Young-type distribution of maxima and minima, without no longer noting the presence of the leading maxima.
In this case, the distance between adjacent minima decreases from the central fringes to the marginal ones, which is consistent with the fact that, in single Airy beams, the front maxima are wider than the rearmost ones.
However, as the tail of beams gets shorter and shorter, such a distance becomes of nearly the same size, as it would correspond to the interference of two Gaussian beams.


\section*{Acknowledgments}
Financial support from the Spanish Research Agency (AEI) and the European
Regional Development Fund (ERDF) (Grant No. PID2021-127781NB-I00) is
acknowledged.



  \bibliographystyle{elsarticle-num} 


\biboptions{sort&compress}

\end{document}